\documentclass[%
 amsmath,amssymb,
preprint,%
]{revtex4-1}

\usepackage{graphicx}
\usepackage{dcolumn}
\usepackage{bm}

\usepackage[utf8]{inputenc}
\usepackage[T1]{fontenc}
\usepackage{mathptmx}
\usepackage{etoolbox}
\usepackage[version=4]{mhchem} 
\usepackage{siunitx} 

\makeatletter
\def\@email#1#2{%
 \endgroup
 \patchcmd{\titleblock@produce}
  {\frontmatter@RRAPformat}
  {\frontmatter@RRAPformat{\produce@RRAP{*#1\href{mailto:#2}{#2}}}\frontmatter@RRAPformat}
  {}{}
}%
\makeatother

\begin{document}


\title
{High-power femtosecond mid-IR source with tunable center frequency and chirp}
\author{Larissa Boie}
    \altaffiliation[Present address: ]{Laboratory for Synchtrotron Radiation and Femtochemistry, Paul Scherrer Institute, Villigen PSI, Switzerland}
    \affiliation{Institute for Quantum Electronics, Physics Department, ETH Zurich, Zurich, Switzerland}

\author{Benjamin H.~Strudwick}%
    \affiliation{SwissFEL, Paul Scherrer Institute, Villigen, Switzerland}

\author{Rafael T.~Winkler}
    \affiliation{Institute for Quantum Electronics, Physics Department, ETH Zurich, Zurich, Switzerland}

\author{Yunpei Deng}
    \affiliation{SwissFEL, Paul Scherrer Institute, Villigen, Switzerland}
    
\author{Steven L.~Johnson}
    \altaffiliation[Corresponding author: ]{johnson@phys.ethz.ch}
    \affiliation{Institute for Quantum Electronics, Physics Department, ETH Zurich, Zurich, Switzerland}
    \affiliation{SwissFEL, Paul Scherrer Institute, Villigen, Switzerland}

\date{\today}

\begin{abstract}
We present an experimental implementation of a chirped mid-infrared (mid-IR) high-power laser source with variable center frequency between \qtyrange{4}{30}{THz} and continuously tunable frequency sweep of up to \qty{20}{\percent} within one pulse, with a pulse duration of \qty{2}{ps}. The peak electric field obtained at \qty{4}{THz} is \qty{1.5}{MV/cm}. We generate the mid-IR light using a difference-frequency generation (DFG) process with two phase-locked, chirped IR pulses. The obtained mid-IR electric field waveform is characterized using electro-optic sampling. We compare our experimental results with the predictions of numerical simulations. The results indicate the potential for efficient driving of vibrational modes into a strongly anharmomic regime, in cases where using Fourier-transform-limited pulses to achieve similar vibrational amplitudes would lead to dielectric breakdown.
\end{abstract}


\maketitle

Although pulsed sources of narrowband radiation in the tera\-hertz range have been available for decades \cite{lee2000,danielson2008}, there has been recently a renewed attention on high-peak-power sources in the \qtyrange{4}{30}{THz} range\cite{liu2016,cartella2017}.
In this frequency range, excitation with narrowband short light pulses can selectively drive particular vibrational modes to extremely high amplitudes, which in some cases can drive materials into new metastable phases via nonlinear vibrational interactions~\cite{foerst2011,mankowsky2014,mankowsky2017,mankowsky2017b,liu2020,subedi2017}. Although similar phenomena have been predicted theoretically in other materials, in many cases triggering these phenomena using Fourier-transform limited pulses require peak electric fields in the mid-IR pulse that are currently difficult to produce or are sufficiently intense to excite competing channels via strongly nonlinear mechanisms~\cite{neugebauer2021}.  

A possible solution for making the driving of nonlinear phononic processes more efficient has been proposed by Itin and Katsnelson \cite{itin2018}: the concept of \textit{capture into resonance}. In cases where light is used to directly drive an infrared-active vibrational mode into a strongly anharmonic regime, the frequency of the mode becomes a function of the vibrational amplitude. The idea of capture into resonance is to apply a small chirp to the mid-IR pulse to match the electromagnetic frequency with the mode frequency, so that less pulse energy is needed to drive the system into a regime where the nonlinear interactions become influential.

\begin{figure*}[htbp]
\centering\includegraphics[width=0.85\textwidth]{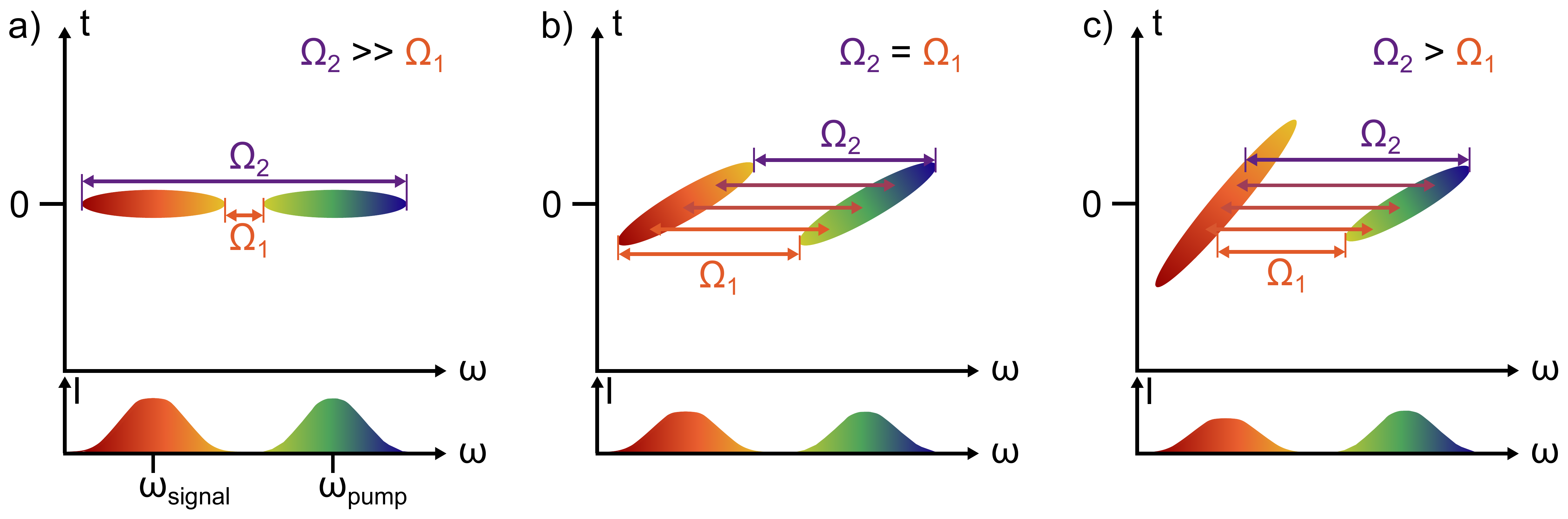}
\caption{Sketches of spectrograms of the DFG process using IR pulses of different chirps in a time-frequency representation. (a) Zero chirp in pump and signal pulses, resulting in a broadband transform-limited mid-infrared pulse. 
(b) Identical chirp applied to both the signal and pump pulses leads to a longer pulse duration and a narrowband (single frequency) idler pulse ($ \Omega_1 = \Omega_2$). (c) Different amounts of chirp applied in both IR pulses gives narrowband mid-infrared pulses with a nonzero chirp from the change in difference frequency with time.}
\label{fig:idea_sketch}
\end{figure*}

Implementing such schemes requires a source of intense mid-IR pulses with tunable frequency and chirp. Free electron lasers (FEL) offer a very high intensity but require a large investment and have non-trivial space requirements \cite{dienst2013}. Interferometric designs making use of pulse beating schemes have become more popular \cite{vicario2020, ji2023}, but are somewhat limited in tunability and can suffer from lower electric field strengths \cite{uchida2015}.   Building on a flexible setup for narrowband mid-IR generation presented by Liu \textit{et al.}~\cite{liu2016}, we present here a carrier-envelope-stable, chirped mid-IR narrowband source with a high degree of tunability. The concept is shown in Fig.~\ref{fig:idea_sketch}.  We employ a difference-frequency generation (DFG) process, where two individually tunable, chirped near-infrared pulses are mixed to obtain a mid-infrared pulse at their difference frequency. By tuning the chirp of one of the near-infrared pulses, the instantaneous value of the difference frequency changes with time, resulting in a mid-infrared pulse with a continuously adjustable chirp.  As demonstrated by Liu \textit{et al.}~\cite{liu2016}, we can also tune the center frequency of the mid-infrared light by changing the temporal delay between the two near-infrared pulses. 

\begin{figure*}[htbp]
\centering\includegraphics[width=0.7\textwidth]{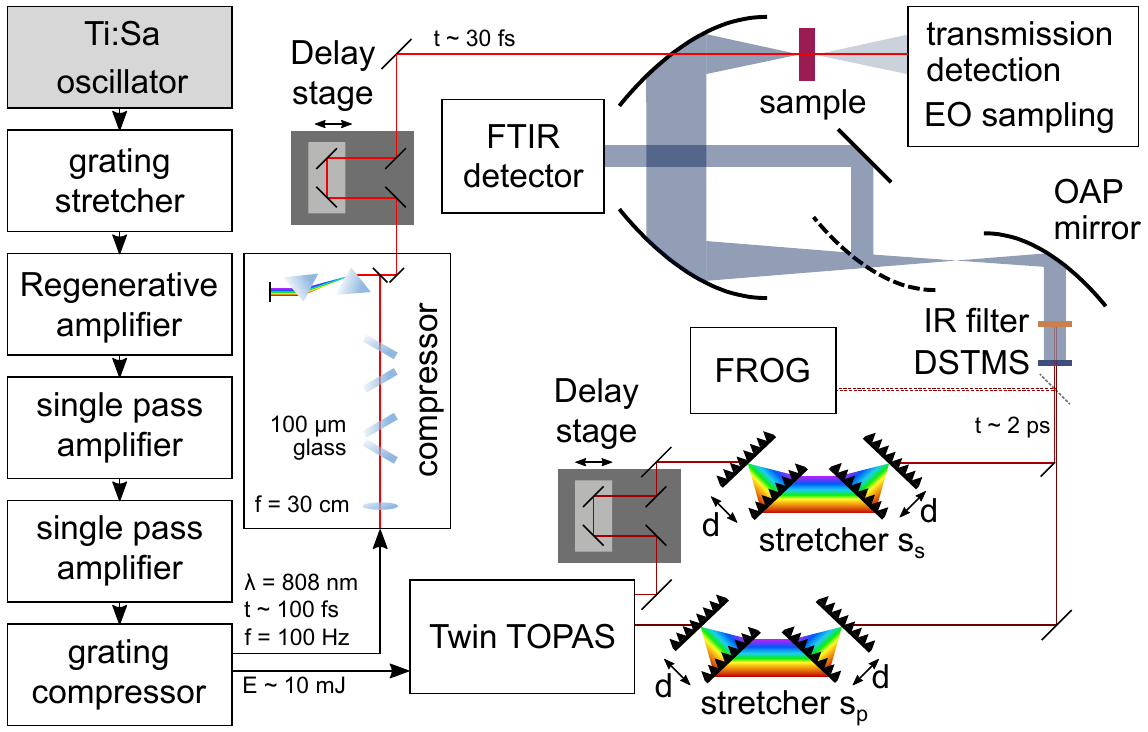}
\caption{Sketch of the experimental setup used to generate chirped mid-IR pulses. The grating distances $d_\text{pump}$ and $d_\text{signal}$ between the transmission gratings are continuously tunable and are set to one value for the pump path ($d_\text{pump}$) and varied for the signal path ($d_\textrm{signal}$) during the experiments. The \ce{GaP} crystal for EO sampling is placed at the sample position. Here ``FROG'' denotes a frequency-resolved optical gating device, ``OAP mirror'' is an off-axis parabolic mirror, and ``FTIR detection'' is a commercial Fourier Transform infrared spectrometer.}
\label{fig:lhx_sketch}
\end{figure*}

To demonstrate the idea of chirped mid-infrared pulses, we use the setup shown in Fig.~\ref{fig:lhx_sketch}.  A high-power amplifier based on a Ti:\ce{Al2O3} laser generates compressed pulses (\qty{100}{Hz}, \qty{808}{nm}, 100 fs, 20 mJ). This output is split into two branches, one for generating the mid-infrared pulses and one for electro-optic sampling of their waveform.

The mid-infrared generation branch begins by driving a pair of commercial optical parametric amplifiers seeded by a common continuum source (Twin TOPAS, LightConversion).  We use the signal outputs of these amplifiers, generating independently tuned pulses within a \qtyrange{1300}{1550}{nm} wavelength range. In the following discussion we will refer to the pulses with higher center frequency as the ``pump'' and the pulses with lower center frequency as the ``signal,'' with reference to their function in the following difference-frequency-generation process.  These two output pulses then pass through two custom-made transmission grating stretchers.

Each transmission grating stretcher consists of four T-966C gratings (LightSmyth Technologies, \sisetup{separate-uncertainty}\qty{0.675 \pm 0.05}{mm} thickness)~\cite{liu2016}. The distances $d_\text{pump}$ and $d_\text{signal}$ between the first and second grating pairs are tuned using a linear micrometer translation stage connected to the mounting plate of the two inner gratings, keeping the outer gratings at a fixed position. 
This distance sets the chirp applied to each pulse~\cite{hecht2017,diels2006}. We keep the chirp of the pump pulse constant, setting the distance within each grating pair to \(d_\textrm{pump} = \qty{3.90}{mm}\). The chirp of the signal pulse is set by tuning the grating distance \(d_\textrm{signal}\) over a range from \qtyrange{2.80}{4.00}{mm} by moving the micrometer stage in steps of \qty{0.10}{mm}. Changing \(d_\textrm{signal}\) leads to a different signal arrival time in the DFG crystal, which we compensate with a delay stage in the signal path.

The signal and pump beams are telescoped to \qty{2}{mm} beam diameter full width half maximum (FWHM), and then directed into a \qty{330}{\micro\meter} thick DSTMS crystal (4-N,N-dimethylamino-4’-N’-methylstilbazolium 2,4,6-trimethylbenzenesulfonate, Rainbow Photonics) where they are mixed collinearly (angular separation < \ang{0.5}) to obtain the difference frequency.  
For our tests we used $\omega_\text{signal} = \qty{210}{THz}$ (\qty{1426}{nm}) and $\omega_\text{pump} = \qty{214}{THz}$ (\qty{1400}{nm}), resulting in a \qty{4}{THz} mid-infrared idler pulse with \qty{1.8}{\micro J} pulse energy from \qty{480}{\micro J} pump pulse energy and \qty{80}{\micro J} signal pulse energy, respectively. 
The beam path in the mid-IR (after the DFG process) is enclosed in a box purged with nitrogen gas for a low humidity environment. 

In order to characterize the resulting mid-IR pulses, they are relayed into a \qty{200}{\micro\meter} thick \ce{GaP} crystal using a sequence of off-axis parabolic (OAP) mirrors.
In a parallel branch, a portion of the compressed output of the TI:Sapphire amplifier (\qty {30}{\micro J}) is further compressed to \qty{30}{fs} using a series of thin glass plates to generate a supercontinuum and then compressing using a pair of prisms~\cite{lu2014,hadrich2016}.  A small fraction of these compressed pulses are then focused into the GaP crystal in the center of the mid-IR beam and used to measure the electric-field waveform of the mid-IR with an electro-optic (EO) sampling scheme~\cite{wu1995,Lu1997}.

\begin{figure*}[htbp]
\centering\includegraphics[width=0.9\textwidth]{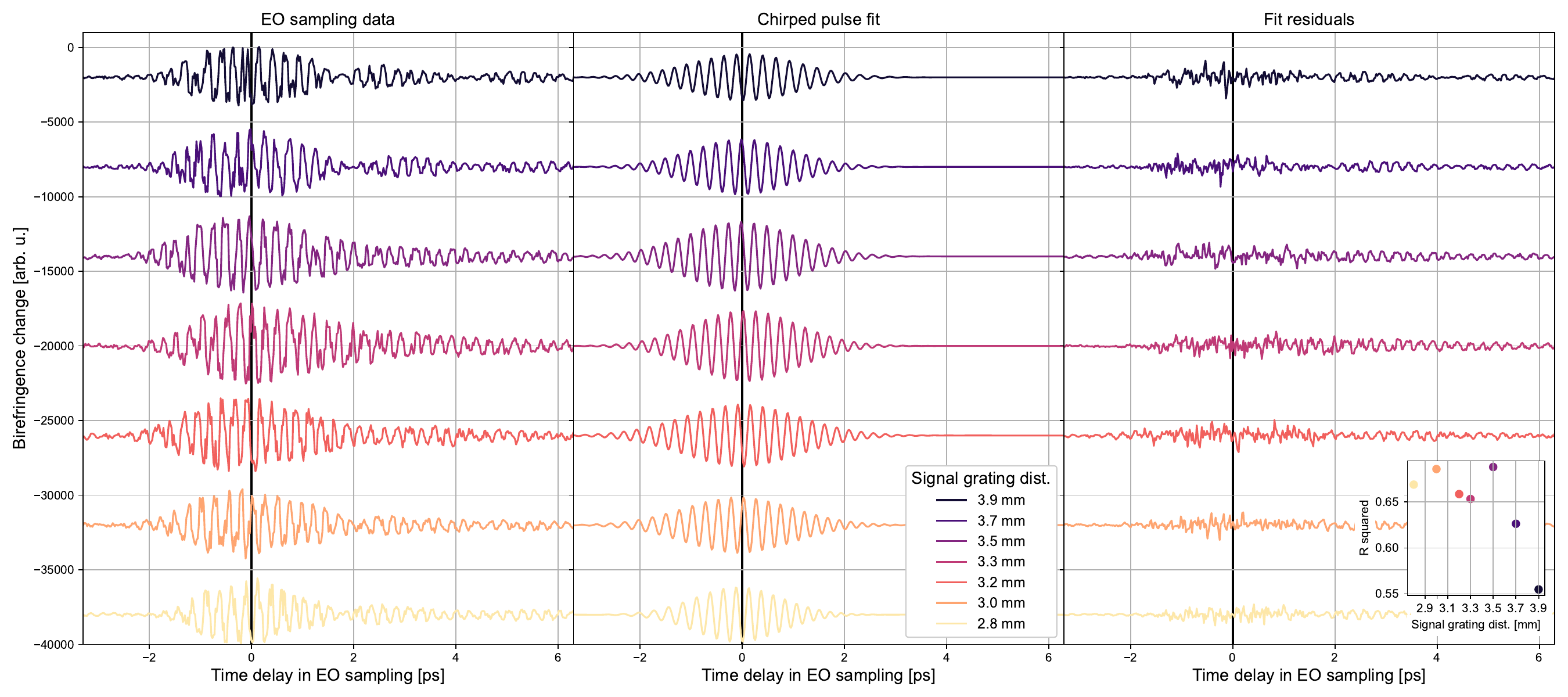}
\caption{Idler electric field measured with electro-optic sampling, with corresponding fit with a chirped electric field pulse (c.f.~Eq.~\ref{eq:chirped_Efield}) and residuals of the fit function (inset shows the $R^2$ value derived from the residuals). The phase of the electric field pulse varies with varying signal grating distance $d_{\text{signal}}$, indicated with the black vertical lines.}
\label{fig:idler_efield_fit}
\end{figure*}

To extract an estimate for the chirp, the idler electric field measured by EO sampling is fit to a model of a Gaussian pulse 
\begin{equation}
    E(t) = A e^{-(t-t_0)^2/(2\sigma^2)} \cdot \sin\left[\omega (t-t_1) + b(t-t_0)^2\right]
\label{eq:chirped_Efield}
\end{equation}
with amplitude $A$, standard deviation $\sigma$, chirp parameter $b$, and time offsets $t_0$ and $t_1$. The electric field and the corresponding fit and residuals are shown in Fig.~\ref{fig:idler_efield_fit} for different grating distances in the stretcher of the signal pulse, where we have also defined the time axis such that $t_0 = 0$ for each pulse.

\begin{figure*}[htbp]
\centering\includegraphics[width=\textwidth]{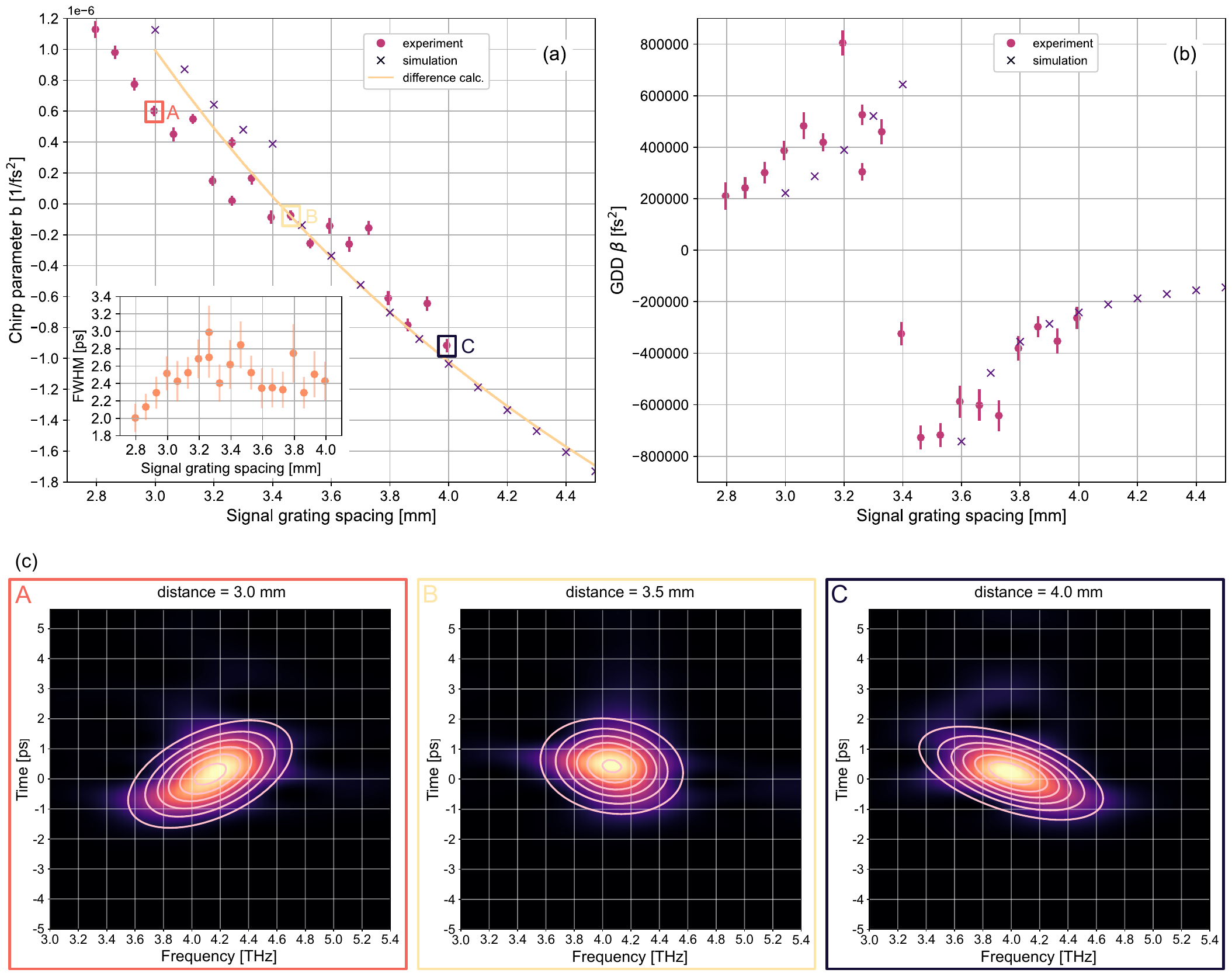}
\caption{\textbf{(a)} Chirp parameter $b$ obtained from a chirped pulse fit of the idler electric field (c.f.~Fig.~\ref{fig:idler_efield_fit}) as a function of signal grating distance. The error bars shown are the asymptotic standard errors estimated by the fit. The dots are derived from the measured electric field pulses, the crosses from the simulation. The corresponding FWHM of the experimental idler electric field is shown in the inset. The continuous line shows the idler chirp as calculated from the difference between the input parameters of pump and signal chirp for the simulation. \textbf{(b)} Group delay dispersion $\beta$ as a function of signal grating distance calculated with Eq.~\ref{eq:GDD} using the chirp values of panel (a). \textbf{(c)} Spectrograms at three selected signal grating distances (A, B, C corresponding to \qtylist{3.0;3.5;4.0}{mm} in panel (a), respectively) showing the two-dimensional representation of the chirp for the experimental data.}
\label{fig:chirp_and_GDD}
\end{figure*}

Figure \ref{fig:chirp_and_GDD}a shows the chirp parameter $b$ as a function of stretcher grating distance $d_\textrm{signal}$. We observe here one of our main results: by tuning the signal grating distance $d_\textrm{signal}$ we continuously change the chirp of the idler pulse. 
For \qty{3.4}{mm} signal grating distance, the idler chirp value is close to zero, indicating a transform-limited narrowband mid-IR pulse without a temporal chirp. For shorter grating distances, the chirp is positive, and for longer grating distances, the chirp is negative. We also note that the slope is larger for shorter grating distances than it is for longer distances. 

The inset of  Fig.~\ref{fig:chirp_and_GDD}a shows the pulse FWHM ($2 \sqrt{2 \ln 2} \sigma$) as a function of $d_\textrm{signal}$. For $d_{\text{signal}}< \qty{3.2}{mm}$, the pulse duration increases with $d_\textrm{signal}$, but levels off for larger values of $d_\textrm{signal}$. This can be explained by considering the relative duration of the pump and signal pulses.  As we increase $d_\textrm{signal}$ we increase the duration of the signal pulse.  For very small values of $d_\textrm{signal}$ the signal pulse is shorter than the pump pulse, and so the resulting idler pulse duration is set by the duration of the signal.  As $d_\textrm{signal}$ increases, the signal becomes longer than the pump pulse.  Here the idler FWHM is limited to the FWHM of the pump.

An alternate parameterization of the deviation of the mid-IR pulse from the Fourier transform limit is the group-delay-dispersion (GDD).  The GDD can be derived from the results of our time-domain fits via~\cite{hong2002}

\begin{equation}
    \beta = \frac{b}{1/(4\sigma^4) + 4 b^2}\label{eq:GDD}
\end{equation}
The resulting $\beta$ values are shown in Fig.~\ref{fig:chirp_and_GDD}b. We observe that while $\beta$ changes rapidly with $d_\textrm{signal}$ near the zero-crossing at 3.2 mm, for larger deviations the magnitude of the GDD rapidly peaks and then slowly relaxes back to smaller values.   

As an alternative method of visualizing the chirp of the resulting idler pulses, we also construct spectrograms of the idler electric field pulse as discussed by Hong \textit{et al.}\cite{hong2002}. We can make use of the phase information obtained during the EO sampling measurement and plot the time-frequency relation within the idler electric field for several signal grating distances $d_\text{signal}$. In Fig.~\ref{fig:chirp_and_GDD}a, three different chirp values are highlighted with capital letters A, B, and C and their corresponding spectrograms are drawn below in Fig.~\ref{fig:chirp_and_GDD}c.
Fitting a 2D Gaussian beam profile to the obtained spectrograms using
\begin{widetext}
\begin{equation}
     f(x,y) = z_0 + A e^{ - (h(x-x_0)^2 + 2 k\cdot(x-x_0)(y-y_0) + l(y-y_0)^2) },\label{eq:2D}
\end{equation}
\end{widetext}
where $z_0$ is the intensity offset, $A$ the amplitude of the pulse, $h = (\cos^2\theta)/(2 \sigma_x^2) + (\sin^2\theta)/(2\sigma_y^2)$, $k = -\sin(2\theta)/(4\sigma_x^2) + \sin(2\theta)/(4\sigma_y^2)$, $l = \sin^2\theta/(2\sigma_x^2) + \cos^2\theta/(2\sigma_y^2)$ with $\theta$ being the the rotation angle of the ellipse and $\sigma_{x,y}$ the widths along each axis, we are able to determine the frequency sweep within the pulse in terms of the slope given by the angle $\theta$ of the fit function, retrieving information on the chirp through a different method than fitting to Eq.~\ref{eq:chirped_Efield}. At $d_\text{signal} = \qty{3.0}{mm}$ (position A), we can clearly see a positive chirp, as the slope indicated by $\theta$ is positive. 
For $d_\text{signal} = \qty{3.5}{mm}$ (B), we see that the Gaussian profile is round, indicating the position where the idler electric field pulse has maximum compression over its duration, thus $b = 0$. At $d_\text{signal} = \qty{4.0}{mm}$ (C), we obtain a negative chirp indicated by the negative slope in the spectrogram. Note that the pulse duration of \qty{2}{ps} is constant over all three configurations, showing that the only parameter varying is indeed the chirp of the pulse. The amount of possible frequency sweep is limited by the initial bandwidth of the IR pulses (in the presented setup approx.~\qty{30}{nm}).

To complement the experimental analysis, we simulated the frequency mixing process using two chirped IR pulses in a material described by the refractive index of DSTMS \cite{jazbinsek2019}, using constant dispersion: $n_\text{pump} = \num{2.1}$, $n_\text{signal} = \num{2.1}$, $n_\text{idler} = \num{2.2}$. The propagation of the electric field pulses through the material is modeled for small spatial steps (discretization of \qty{1}{\micro\meter}) where the transmission is set to \num{0.995} at each step, solving the coupled wave equations\cite{boyd2003} numerically using the split-step method. The thickness of the non-linear material was set to \qty{0.2}{mm}, which is slightly thinner than the real DSTMS crystal used. Also, the pulse energies are set to lower values ($E_{\text{pump}} = \qty{0.1}{mJ}$ and $E_{\text{signal}} = \qty{0.01}{mJ}$ for a beam diameter of \qty{6.0}{mm}).  Figure \ref{fig:chirp_and_GDD}a shows the chirp parameter $b$ for each electric field pulse obtained from the simulation, using a fit to Eq.~\ref{eq:chirped_Efield}, while Fig.~\ref{fig:chirp_and_GDD}b shows the corresponding group delay dispersion $\beta$ (purple crosses). 

We observe that the simulated DFG process is more efficient in the simulations: \qty{0.9}{\percent} energy conversion efficiency, versus \qty{0.3}{\percent} in the experiment.  This may be due to the lack of consideration of absorption losses and scattering effects in the simulation. We also observe a small systematic difference in the value of $d_\textrm{signal}$ between the results of the simulation and the experiment for smaller values of $d_\textrm{signal}$, of about \qty{0.2}{mm} in magnitude.  This may be at least partly caused by experimental inaccuracies in determining the absolute position of $d_\textrm{signal}$.

With the simulation, we are able to extend the amount of applied chirp in each IR pulse, studying extreme regimes where the experimental conditions are limiting the DFG process. We can identify an asymptotic behaviour, which yields information on the pump and signal GDD values that are experimentally difficult to measure and require an additional measurement step. For very large signal grating distances $d_{\text{signal}}$, $\beta_\text{signal} \gg \beta_\text{pump}$ and $\beta_\text{idler}$ approaches the $\beta_\text{pump}$ value. For very short $d_{\text{signal}}$, $\beta_\text{idler}$ approaches zero. For the shown simulation, the pump GDD is constant with $\beta = \qty{-70219}{fs^2}$, calculated using the grating equation\cite{hecht2017,diels2006}. The idler GDD for a signal grating distance of \qty{6.0}{mm} obtained from the simulation is \qty{-80018}{fs^2}. As the maximum chirp obtained during the simulation is \qty{-742916}{fs^2}, the idler GDD has approached the pump GDD within \qty{1}{\percent} of the maximum applied chirp. 

\begin{figure*}[htbp]
\centering\includegraphics[width=0.9\textwidth]{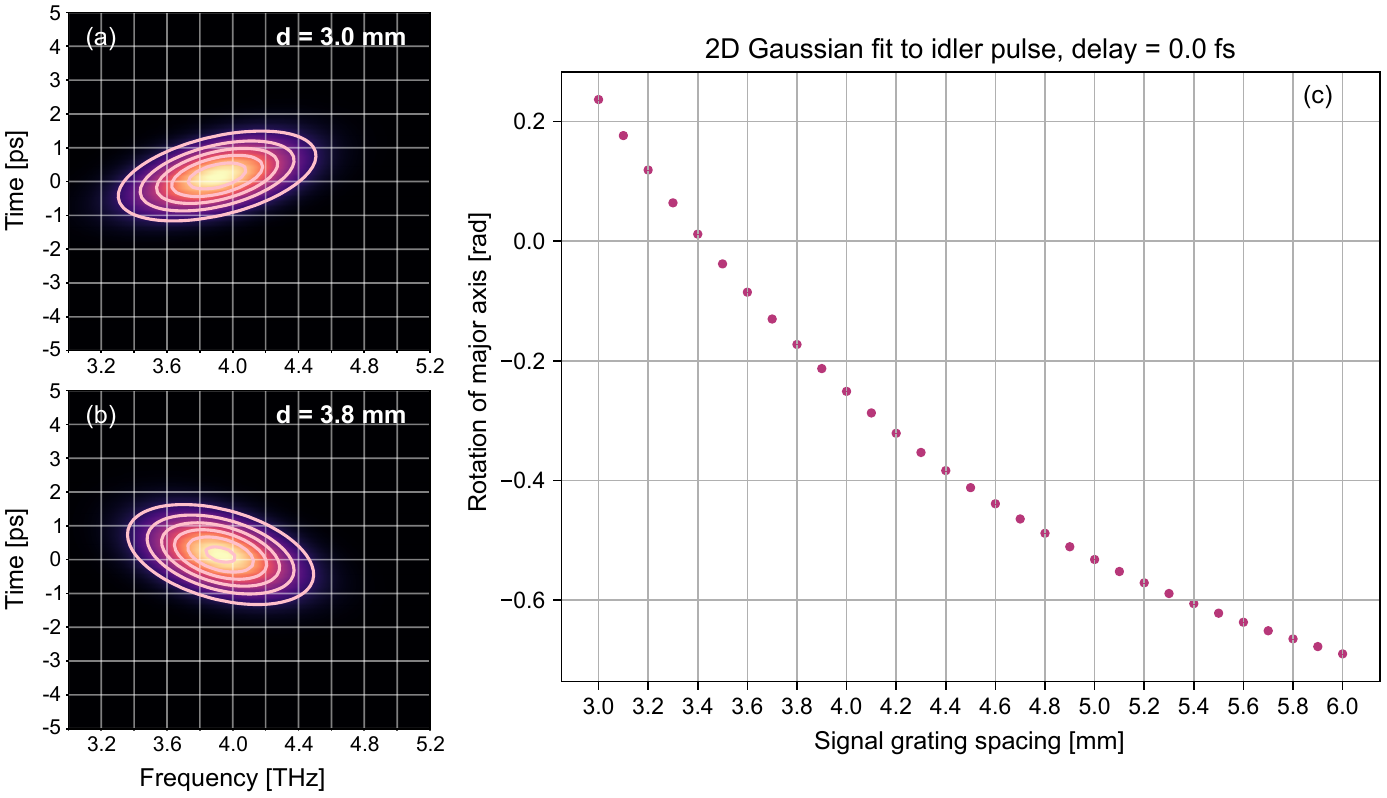}
\caption{Two dimensional analysis of the simulation results shown as spectrograms. \textbf{(a)}: Mid-IR pulse with positive chirp applied, grating distance \qty{3.0}{mm}. \textbf{(b)}: Mid-IR pulse with negative chirp applied, grating distance \qty{3.8}{mm}. \textbf{(c)}: Rotation of the major axis as obtained from a 2D Gaussian pulse fit as indicated with the contour lines in (a) and (b). The rotation is a measure for the amount of applied chirp. The curvature and the signal grating distance for the zero chirp case match the one dimensional analysis well.}
\label{fig:sim_2D_analysis}
\end{figure*}

Similar to the analysis of the experimental data, we construct spectrograms of the simulation results and fit them to Eq.~\ref{eq:2D}.  This analysis is summarized in Fig.~\ref{fig:sim_2D_analysis}. Both the bandwidth and the slope agree well with the experimental results shown in Fig.~\ref{fig:chirp_and_GDD}c. The curve of the experimental chirp parameter $b$ matches the curve of the rotation $\theta$ from the 2D Gaussian fits to the spectrograms of the simulated results. 

In summary, we have presented a flexible chirped mid-IR source with pulse energies up to \qty{1.8}{\micro J} delivering \qty{1.5}{MV/cm} electric field strength using a DFG process between two chirped IR pulses in the organic crystal DSTMS that are each shaped with a custom-built transmission grating stretcher. Tuning the distance within a grating pair in the signal stretcher, we are able to adjust the amount of chirp obtained in the mid-IR pulse. We have extensively characterized the chirp tunability, showing the maximum and minimum applied group delay dispersion (GDD) for a certain pump and signal chirp combination, and estimated the applied GDD of the pump pulse from the asymptotic behaviour of the GDD as a function of signal grating distance. The experiments match the simulations presented here, allowing for a precise tuning of the chirp parameters and IR wavelengths. Further pulse shaping techniques for the IR pulses that increase the bandwidths can allow for larger frequency changes, as the maximum possible frequency sweep is defined by the initial bandwidths in the pump and signal pulses.

Our results complement efforts to provide large tunability and flexible excitation over the terahertz and mid-IR spectrum for nonlinear applications. By mixing the outputs of two separately tunable OPAs as opposed to mixing the signal and idler of a single OPA, we obtain both carrier-envelope phase stability and a large tunability while keeping the laser output constant. Experiments using intense mid-IR sources for pump and probe schemes have become more common \cite{cartella2018} and narrowband pulses in this frequency regime enable selective pumping \cite{henstridge2022,basini2024}. In comparison to schemes using one laser pulse in combination with either an interferometric setup \cite{vicario2020} or etalon-based chirped-pulse beating \cite{ji2023}, our approach does not require the replacement of optics like an etalon to change the pulse characteristics. This increases the control over the pulse envelope and enables full tunability via translation stages. Recent applications of a similar design show the versatility of our approach, employing a DFG scheme in another organic crystal (DAST) for THz electric field-induced second harmonic generation in microscopy applications \cite{lin2023}.

\section*{Funding}
This research was supported by the NCCR MUST, funded by the Swiss National Science Foundation. We also acknowledge funding by the Swiss National Science Foundation through Grant 192337.

\section*{Author Declarations}
\subsection*{Disclosures}
\noindent The authors have no conflicts to disclose.

\subsection*{Author Contributions}
\noindent\textbf{L.~Boie:} Data curation (lead), formal analysis (equal), conceptualization (support), methodology (equal), investigation (equal), project administration (support), software (support), validation (equal), visualization (lead), writing -- original draft (equal)
\textbf{B.~H.~Strudwick:} Formal analysis (equal), investigation (equal), methodology (equal), software (equal), validation (equal), visualization (suport), writing -- original draft (support)
\textbf{R.~Winkler:} Methodology (equal), investigation (equal), formal analysis (support), software (equal), visualization (support), writing -- original draft (support)
\textbf{Y.~Deng:} Data curation (support), methodology (support), investigation (equal), validation (equal), conceptualization (support), writing -- original draft (support)
\textbf{S.~L.~Johnson:} Conceptualization (lead), funding acquisition (lead), methodology (lead), project administration (lead), supervision (lead), writing -- original draft (equal), validation (support)

\section*{Data Availability Statement}
The data that support the findings of this study are openly available in the ETHZ Research Collection at \url{https://doi.org/10.3929/ethz-b-000584926}. \\

\bibliography{litlist}

\begin{thebibliography}{29}%
\makeatletter
\providecommand \@ifxundefined [1]{%
 \@ifx{#1\undefined}
}%
\providecommand \@ifnum [1]{%
 \ifnum #1\expandafter \@firstoftwo
 \else \expandafter \@secondoftwo
 \fi
}%
\providecommand \@ifx [1]{%
 \ifx #1\expandafter \@firstoftwo
 \else \expandafter \@secondoftwo
 \fi
}%
\providecommand \natexlab [1]{#1}%
\providecommand \enquote  [1]{``#1''}%
\providecommand \bibnamefont  [1]{#1}%
\providecommand \bibfnamefont [1]{#1}%
\providecommand \citenamefont [1]{#1}%
\providecommand \href@noop [0]{\@secondoftwo}%
\providecommand \href [0]{\begingroup \@sanitize@url \@href}%
\providecommand \@href[1]{\@@startlink{#1}\@@href}%
\providecommand \@@href[1]{\endgroup#1\@@endlink}%
\providecommand \@sanitize@url [0]{\catcode `\\12\catcode `\$12\catcode `\&12\catcode `\#12\catcode `\^12\catcode `\_12\catcode `\%12\relax}%
\providecommand \@@startlink[1]{}%
\providecommand \@@endlink[0]{}%
\providecommand \url  [0]{\begingroup\@sanitize@url \@url }%
\providecommand \@url [1]{\endgroup\@href {#1}{\urlprefix }}%
\providecommand \urlprefix  [0]{URL }%
\providecommand \Eprint [0]{\href }%
\providecommand \doibase [0]{http://dx.doi.org/}%
\providecommand \selectlanguage [0]{\@gobble}%
\providecommand \bibinfo  [0]{\@secondoftwo}%
\providecommand \bibfield  [0]{\@secondoftwo}%
\providecommand \translation [1]{[#1]}%
\providecommand \BibitemOpen [0]{}%
\providecommand \bibitemStop [0]{}%
\providecommand \bibitemNoStop [0]{.\EOS\space}%
\providecommand \EOS [0]{\spacefactor3000\relax}%
\providecommand \BibitemShut  [1]{\csname bibitem#1\endcsname}%
\let\auto@bib@innerbib\@empty
\bibitem [{\citenamefont {Lee}\ \emph {et~al.}(2000)\citenamefont {Lee}, \citenamefont {Meade}, \citenamefont {DeCamp}, \citenamefont {Norris},\ and\ \citenamefont {Galvanauskas}}]{lee2000}%
  \BibitemOpen
  \bibfield  {author} {\bibinfo {author} {\bibfnamefont {Y.-S.}\ \bibnamefont {Lee}}, \bibinfo {author} {\bibfnamefont {T.}~\bibnamefont {Meade}}, \bibinfo {author} {\bibfnamefont {M.}~\bibnamefont {DeCamp}}, \bibinfo {author} {\bibfnamefont {T.~B.}\ \bibnamefont {Norris}}, \ and\ \bibinfo {author} {\bibfnamefont {A.}~\bibnamefont {Galvanauskas}},\ }\href {\doibase 10.1063/1.1290046} {\bibfield  {journal} {\bibinfo  {journal} {Appl. Phys. Lett.}\ }\textbf {\bibinfo {volume} {77}},\ \bibinfo {pages} {1244} (\bibinfo {year} {2000})}\BibitemShut {NoStop}%
\bibitem [{\citenamefont {Danielson}\ \emph {et~al.}(2008)\citenamefont {Danielson}, \citenamefont {Jameson}, \citenamefont {Tomaino}, \citenamefont {Hui}, \citenamefont {Wetzel}, \citenamefont {Lee},\ and\ \citenamefont {Vodopyanov}}]{danielson2008}%
  \BibitemOpen
  \bibfield  {author} {\bibinfo {author} {\bibfnamefont {J.~R.}\ \bibnamefont {Danielson}}, \bibinfo {author} {\bibfnamefont {A.~D.}\ \bibnamefont {Jameson}}, \bibinfo {author} {\bibfnamefont {J.~L.}\ \bibnamefont {Tomaino}}, \bibinfo {author} {\bibfnamefont {H.}~\bibnamefont {Hui}}, \bibinfo {author} {\bibfnamefont {J.~D.}\ \bibnamefont {Wetzel}}, \bibinfo {author} {\bibfnamefont {Y.-S.}\ \bibnamefont {Lee}}, \ and\ \bibinfo {author} {\bibfnamefont {K.~L.}\ \bibnamefont {Vodopyanov}},\ }\href {\doibase 10.1063/1.2959846} {\bibfield  {journal} {\bibinfo  {journal} {J. Appl. Phys.}\ }\textbf {\bibinfo {volume} {104}} (\bibinfo {year} {2008}),\ 10.1063/1.2959846}\BibitemShut {NoStop}%
\bibitem [{\citenamefont {Liu}\ \emph {et~al.}(2017)\citenamefont {Liu}, \citenamefont {Bromberger}, \citenamefont {Cartella}, \citenamefont {Gebert}, \citenamefont {F{\"o}rst},\ and\ \citenamefont {Cavalleri}}]{liu2016}%
  \BibitemOpen
  \bibfield  {author} {\bibinfo {author} {\bibfnamefont {B.}~\bibnamefont {Liu}}, \bibinfo {author} {\bibfnamefont {H.}~\bibnamefont {Bromberger}}, \bibinfo {author} {\bibfnamefont {A.}~\bibnamefont {Cartella}}, \bibinfo {author} {\bibfnamefont {T.}~\bibnamefont {Gebert}}, \bibinfo {author} {\bibfnamefont {M.}~\bibnamefont {F{\"o}rst}}, \ and\ \bibinfo {author} {\bibfnamefont {A.}~\bibnamefont {Cavalleri}},\ }\href {\doibase 10.1364/ol.42.000129} {\bibfield  {journal} {\bibinfo  {journal} {Opt. Lett.}\ }\textbf {\bibinfo {volume} {42}},\ \bibinfo {pages} {129} (\bibinfo {year} {2017})}\BibitemShut {NoStop}%
\bibitem [{\citenamefont {Cartella}\ \emph {et~al.}(2017)\citenamefont {Cartella}, \citenamefont {Nova}, \citenamefont {Oriana}, \citenamefont {Cerullo}, \citenamefont {F\"{o}rst}, \citenamefont {Manzoni},\ and\ \citenamefont {Cavalleri}}]{cartella2017}%
  \BibitemOpen
  \bibfield  {author} {\bibinfo {author} {\bibfnamefont {A.}~\bibnamefont {Cartella}}, \bibinfo {author} {\bibfnamefont {T.~F.}\ \bibnamefont {Nova}}, \bibinfo {author} {\bibfnamefont {A.}~\bibnamefont {Oriana}}, \bibinfo {author} {\bibfnamefont {G.}~\bibnamefont {Cerullo}}, \bibinfo {author} {\bibfnamefont {M.}~\bibnamefont {F\"{o}rst}}, \bibinfo {author} {\bibfnamefont {C.}~\bibnamefont {Manzoni}}, \ and\ \bibinfo {author} {\bibfnamefont {A.}~\bibnamefont {Cavalleri}},\ }\href {\doibase 10.1364/ol.42.000663} {\bibfield  {journal} {\bibinfo  {journal} {Opt Lett}\ }\textbf {\bibinfo {volume} {42}},\ \bibinfo {pages} {663} (\bibinfo {year} {2017})}\BibitemShut {NoStop}%
\bibitem [{\citenamefont {F\"{o}rst}\ \emph {et~al.}(2011)\citenamefont {F\"{o}rst}, \citenamefont {Manzoni}, \citenamefont {Kaiser}, \citenamefont {Tomioka}, \citenamefont {Tokura}, \citenamefont {Merlin},\ and\ \citenamefont {Cavalleri}}]{foerst2011}%
  \BibitemOpen
  \bibfield  {author} {\bibinfo {author} {\bibfnamefont {M.}~\bibnamefont {F\"{o}rst}}, \bibinfo {author} {\bibfnamefont {C.}~\bibnamefont {Manzoni}}, \bibinfo {author} {\bibfnamefont {S.}~\bibnamefont {Kaiser}}, \bibinfo {author} {\bibfnamefont {Y.}~\bibnamefont {Tomioka}}, \bibinfo {author} {\bibfnamefont {Y.}~\bibnamefont {Tokura}}, \bibinfo {author} {\bibfnamefont {R.}~\bibnamefont {Merlin}}, \ and\ \bibinfo {author} {\bibfnamefont {A.}~\bibnamefont {Cavalleri}},\ }\href {\doibase 10.1038/nphys2055} {\bibfield  {journal} {\bibinfo  {journal} {Nat Phys}\ }\textbf {\bibinfo {volume} {7}},\ \bibinfo {pages} {854} (\bibinfo {year} {2011})}\BibitemShut {NoStop}%
\bibitem [{\citenamefont {Mankowsky}\ \emph {et~al.}(2014)\citenamefont {Mankowsky}, \citenamefont {Subedi}, \citenamefont {F\"{o}rst}, \citenamefont {Mariager}, \citenamefont {Chollet}, \citenamefont {Lemke}, \citenamefont {Robinson}, \citenamefont {Glownia}, \citenamefont {Minitti}, \citenamefont {Frano}, \citenamefont {Fechner}, \citenamefont {Spaldin}, \citenamefont {Loew}, \citenamefont {Keimer}, \citenamefont {Georges},\ and\ \citenamefont {Cavalleri}}]{mankowsky2014}%
  \BibitemOpen
  \bibfield  {author} {\bibinfo {author} {\bibfnamefont {R.}~\bibnamefont {Mankowsky}}, \bibinfo {author} {\bibfnamefont {A.}~\bibnamefont {Subedi}}, \bibinfo {author} {\bibfnamefont {M.}~\bibnamefont {F\"{o}rst}}, \bibinfo {author} {\bibfnamefont {S.~O.}\ \bibnamefont {Mariager}}, \bibinfo {author} {\bibfnamefont {M.}~\bibnamefont {Chollet}}, \bibinfo {author} {\bibfnamefont {H.~T.}\ \bibnamefont {Lemke}}, \bibinfo {author} {\bibfnamefont {J.~S.}\ \bibnamefont {Robinson}}, \bibinfo {author} {\bibfnamefont {J.~M.}\ \bibnamefont {Glownia}}, \bibinfo {author} {\bibfnamefont {M.~P.}\ \bibnamefont {Minitti}}, \bibinfo {author} {\bibfnamefont {A.}~\bibnamefont {Frano}}, \bibinfo {author} {\bibfnamefont {M.}~\bibnamefont {Fechner}}, \bibinfo {author} {\bibfnamefont {N.~A.}\ \bibnamefont {Spaldin}}, \bibinfo {author} {\bibfnamefont {T.}~\bibnamefont {Loew}}, \bibinfo {author} {\bibfnamefont {B.}~\bibnamefont {Keimer}}, \bibinfo {author} {\bibfnamefont {A.}~\bibnamefont {Georges}}, \ and\ \bibinfo {author}
  {\bibfnamefont {A.}~\bibnamefont {Cavalleri}},\ }\href {\doibase 10.1038/nature13875} {\bibfield  {journal} {\bibinfo  {journal} {Nature}\ }\textbf {\bibinfo {volume} {516}},\ \bibinfo {pages} {71} (\bibinfo {year} {2014})}\BibitemShut {NoStop}%
\bibitem [{\citenamefont {Mankowsky}\ \emph {et~al.}(2017{\natexlab{a}})\citenamefont {Mankowsky}, \citenamefont {von Hoegen}, \citenamefont {F\"{o}rst},\ and\ \citenamefont {Cavalleri}}]{mankowsky2017}%
  \BibitemOpen
  \bibfield  {author} {\bibinfo {author} {\bibfnamefont {R.}~\bibnamefont {Mankowsky}}, \bibinfo {author} {\bibfnamefont {A.}~\bibnamefont {von Hoegen}}, \bibinfo {author} {\bibfnamefont {M.}~\bibnamefont {F\"{o}rst}}, \ and\ \bibinfo {author} {\bibfnamefont {A.}~\bibnamefont {Cavalleri}},\ }\href {\doibase 10.1103/physrevlett.118.197601} {\bibfield  {journal} {\bibinfo  {journal} {Phys. Rev. Lett.}\ }\textbf {\bibinfo {volume} {118}} (\bibinfo {year} {2017}{\natexlab{a}}),\ 10.1103/physrevlett.118.197601}\BibitemShut {NoStop}%
\bibitem [{\citenamefont {Mankowsky}\ \emph {et~al.}(2017{\natexlab{b}})\citenamefont {Mankowsky}, \citenamefont {Liu}, \citenamefont {Rajasekaran}, \citenamefont {Liu}, \citenamefont {Mou}, \citenamefont {Zhou}, \citenamefont {Merlin}, \citenamefont {F\"{o}rst},\ and\ \citenamefont {Cavalleri}}]{mankowsky2017b}%
  \BibitemOpen
  \bibfield  {author} {\bibinfo {author} {\bibfnamefont {R.}~\bibnamefont {Mankowsky}}, \bibinfo {author} {\bibfnamefont {B.}~\bibnamefont {Liu}}, \bibinfo {author} {\bibfnamefont {S.}~\bibnamefont {Rajasekaran}}, \bibinfo {author} {\bibfnamefont {H.~Y.}\ \bibnamefont {Liu}}, \bibinfo {author} {\bibfnamefont {D.}~\bibnamefont {Mou}}, \bibinfo {author} {\bibfnamefont {X.~J.}\ \bibnamefont {Zhou}}, \bibinfo {author} {\bibfnamefont {R.}~\bibnamefont {Merlin}}, \bibinfo {author} {\bibfnamefont {M.}~\bibnamefont {F\"{o}rst}}, \ and\ \bibinfo {author} {\bibfnamefont {A.}~\bibnamefont {Cavalleri}},\ }\href {\doibase 10.1103/PhysRevLett.118.116402} {\bibfield  {journal} {\bibinfo  {journal} {Phys. Rev. Lett.}\ }\textbf {\bibinfo {volume} {118}} (\bibinfo {year} {2017}{\natexlab{b}}),\ 10.1103/PhysRevLett.118.116402}\BibitemShut {NoStop}%
\bibitem [{\citenamefont {Liu}\ \emph {et~al.}(2020)\citenamefont {Liu}, \citenamefont {F\"{o}rst}, \citenamefont {Fechner}, \citenamefont {Nicoletti}, \citenamefont {Porras}, \citenamefont {Loew}, \citenamefont {Keimer},\ and\ \citenamefont {Cavalleri}}]{liu2020}%
  \BibitemOpen
  \bibfield  {author} {\bibinfo {author} {\bibfnamefont {B.}~\bibnamefont {Liu}}, \bibinfo {author} {\bibfnamefont {M.}~\bibnamefont {F\"{o}rst}}, \bibinfo {author} {\bibfnamefont {M.}~\bibnamefont {Fechner}}, \bibinfo {author} {\bibfnamefont {D.}~\bibnamefont {Nicoletti}}, \bibinfo {author} {\bibfnamefont {J.}~\bibnamefont {Porras}}, \bibinfo {author} {\bibfnamefont {T.}~\bibnamefont {Loew}}, \bibinfo {author} {\bibfnamefont {B.}~\bibnamefont {Keimer}}, \ and\ \bibinfo {author} {\bibfnamefont {A.}~\bibnamefont {Cavalleri}},\ }\href {\doibase 10.1103/PhysRevX.10.011053} {\bibfield  {journal} {\bibinfo  {journal} {Phys. Rev. X}\ }\textbf {\bibinfo {volume} {10}},\ \bibinfo {pages} {011053} (\bibinfo {year} {2020})}\BibitemShut {NoStop}%
\bibitem [{\citenamefont {Subedi}(2017)}]{subedi2017}%
  \BibitemOpen
  \bibfield  {author} {\bibinfo {author} {\bibfnamefont {A.}~\bibnamefont {Subedi}},\ }\href {\doibase 10.1103/physrevb.95.134113} {\bibfield  {journal} {\bibinfo  {journal} {Phys. Rev. B}\ }\textbf {\bibinfo {volume} {95}} (\bibinfo {year} {2017}),\ 10.1103/physrevb.95.134113}\BibitemShut {NoStop}%
\bibitem [{\citenamefont {Neugebauer}\ \emph {et~al.}(2021)\citenamefont {Neugebauer}, \citenamefont {Juraschek}, \citenamefont {Savoini}, \citenamefont {Engeler}, \citenamefont {Boie}, \citenamefont {Abreu}, \citenamefont {Spaldin},\ and\ \citenamefont {Johnson}}]{neugebauer2021}%
  \BibitemOpen
  \bibfield  {author} {\bibinfo {author} {\bibfnamefont {M.~J.}\ \bibnamefont {Neugebauer}}, \bibinfo {author} {\bibfnamefont {D.~M.}\ \bibnamefont {Juraschek}}, \bibinfo {author} {\bibfnamefont {M.}~\bibnamefont {Savoini}}, \bibinfo {author} {\bibfnamefont {P.}~\bibnamefont {Engeler}}, \bibinfo {author} {\bibfnamefont {L.}~\bibnamefont {Boie}}, \bibinfo {author} {\bibfnamefont {E.}~\bibnamefont {Abreu}}, \bibinfo {author} {\bibfnamefont {N.~A.}\ \bibnamefont {Spaldin}}, \ and\ \bibinfo {author} {\bibfnamefont {S.~L.}\ \bibnamefont {Johnson}},\ }\href {\doibase 10.1103/PhysRevResearch.3.013126} {\bibfield  {journal} {\bibinfo  {journal} {Phys. Rev. Research}\ }\textbf {\bibinfo {volume} {3}},\ \bibinfo {pages} {013126} (\bibinfo {year} {2021})}\BibitemShut {NoStop}%
\bibitem [{\citenamefont {Itin}\ and\ \citenamefont {Katsnelson}(2018)}]{itin2018}%
  \BibitemOpen
  \bibfield  {author} {\bibinfo {author} {\bibfnamefont {A.~P.}\ \bibnamefont {Itin}}\ and\ \bibinfo {author} {\bibfnamefont {M.~I.}\ \bibnamefont {Katsnelson}},\ }\href {\doibase 10.1103/PhysRevB.97.184304} {\bibfield  {journal} {\bibinfo  {journal} {Phys. Rev. B}\ }\textbf {\bibinfo {volume} {97}},\ \bibinfo {pages} {184304} (\bibinfo {year} {2018})}\BibitemShut {NoStop}%
\bibitem [{\citenamefont {Dienst}\ \emph {et~al.}(2013)\citenamefont {Dienst}, \citenamefont {Casandruc}, \citenamefont {Fausti}, \citenamefont {Zhang}, \citenamefont {Eckstein}, \citenamefont {Hoffmann}, \citenamefont {Khanna}, \citenamefont {Dean}, \citenamefont {Gensch}, \citenamefont {Winnerl}, \citenamefont {Seidel}, \citenamefont {Pyon}, \citenamefont {Takayama}, \citenamefont {Takagi},\ and\ \citenamefont {Cavalleri}}]{dienst2013}%
  \BibitemOpen
  \bibfield  {author} {\bibinfo {author} {\bibfnamefont {A.}~\bibnamefont {Dienst}}, \bibinfo {author} {\bibfnamefont {E.}~\bibnamefont {Casandruc}}, \bibinfo {author} {\bibfnamefont {D.}~\bibnamefont {Fausti}}, \bibinfo {author} {\bibfnamefont {L.}~\bibnamefont {Zhang}}, \bibinfo {author} {\bibfnamefont {M.}~\bibnamefont {Eckstein}}, \bibinfo {author} {\bibfnamefont {M.}~\bibnamefont {Hoffmann}}, \bibinfo {author} {\bibfnamefont {V.}~\bibnamefont {Khanna}}, \bibinfo {author} {\bibfnamefont {N.}~\bibnamefont {Dean}}, \bibinfo {author} {\bibfnamefont {M.}~\bibnamefont {Gensch}}, \bibinfo {author} {\bibfnamefont {S.}~\bibnamefont {Winnerl}}, \bibinfo {author} {\bibfnamefont {W.}~\bibnamefont {Seidel}}, \bibinfo {author} {\bibfnamefont {S.}~\bibnamefont {Pyon}}, \bibinfo {author} {\bibfnamefont {T.}~\bibnamefont {Takayama}}, \bibinfo {author} {\bibfnamefont {H.}~\bibnamefont {Takagi}}, \ and\ \bibinfo {author} {\bibfnamefont {A.}~\bibnamefont {Cavalleri}},\ }\href {\doibase 10.1038/nmat3580} {\bibfield
  {journal} {\bibinfo  {journal} {Nature Materials}\ }\textbf {\bibinfo {volume} {12}},\ \bibinfo {pages} {535} (\bibinfo {year} {2013})}\BibitemShut {NoStop}%
\bibitem [{\citenamefont {Vicario}\ \emph {et~al.}(2020)\citenamefont {Vicario}, \citenamefont {Trisorio}, \citenamefont {Allenspach}, \citenamefont {R\"{u}egg},\ and\ \citenamefont {Giorgianni}}]{vicario2020}%
  \BibitemOpen
  \bibfield  {author} {\bibinfo {author} {\bibfnamefont {C.}~\bibnamefont {Vicario}}, \bibinfo {author} {\bibfnamefont {A.}~\bibnamefont {Trisorio}}, \bibinfo {author} {\bibfnamefont {S.}~\bibnamefont {Allenspach}}, \bibinfo {author} {\bibfnamefont {C.}~\bibnamefont {R\"{u}egg}}, \ and\ \bibinfo {author} {\bibfnamefont {F.}~\bibnamefont {Giorgianni}},\ }\href {\doibase 10.1063/5.0015612} {\bibfield  {journal} {\bibinfo  {journal} {Appl Phys Lett}\ }\textbf {\bibinfo {volume} {117}},\ \bibinfo {pages} {101101} (\bibinfo {year} {2020})}\BibitemShut {NoStop}%
\bibitem [{\citenamefont {Ji}\ \emph {et~al.}(2023)\citenamefont {Ji}, \citenamefont {Hibberd}, \citenamefont {Lin}, \citenamefont {Walsh}, \citenamefont {Thomson}, \citenamefont {Nutter},\ and\ \citenamefont {Graham}}]{ji2023}%
  \BibitemOpen
  \bibfield  {author} {\bibinfo {author} {\bibfnamefont {R.}~\bibnamefont {Ji}}, \bibinfo {author} {\bibfnamefont {M.~T.}\ \bibnamefont {Hibberd}}, \bibinfo {author} {\bibfnamefont {C.-H.}\ \bibnamefont {Lin}}, \bibinfo {author} {\bibfnamefont {D.~A.}\ \bibnamefont {Walsh}}, \bibinfo {author} {\bibfnamefont {T.}~\bibnamefont {Thomson}}, \bibinfo {author} {\bibfnamefont {P.~W.}\ \bibnamefont {Nutter}}, \ and\ \bibinfo {author} {\bibfnamefont {D.~M.}\ \bibnamefont {Graham}},\ }\href {\doibase 10.1063/5.0176314} {\bibfield  {journal} {\bibinfo  {journal} {Appl. Phys. Lett.}\ }\textbf {\bibinfo {volume} {123}} (\bibinfo {year} {2023}),\ 10.1063/5.0176314}\BibitemShut {NoStop}%
\bibitem [{\citenamefont {Uchida}\ \emph {et~al.}(2015)\citenamefont {Uchida}, \citenamefont {Hirori}, \citenamefont {Aoki}, \citenamefont {Wolpert}, \citenamefont {Tamaya}, \citenamefont {Tanaka}, \citenamefont {Mochizuki}, \citenamefont {Kim}, \citenamefont {Yoshita}, \citenamefont {Akiyama}, \citenamefont {Pfeiffer},\ and\ \citenamefont {West}}]{uchida2015}%
  \BibitemOpen
  \bibfield  {author} {\bibinfo {author} {\bibfnamefont {K.}~\bibnamefont {Uchida}}, \bibinfo {author} {\bibfnamefont {H.}~\bibnamefont {Hirori}}, \bibinfo {author} {\bibfnamefont {T.}~\bibnamefont {Aoki}}, \bibinfo {author} {\bibfnamefont {C.}~\bibnamefont {Wolpert}}, \bibinfo {author} {\bibfnamefont {T.}~\bibnamefont {Tamaya}}, \bibinfo {author} {\bibfnamefont {K.}~\bibnamefont {Tanaka}}, \bibinfo {author} {\bibfnamefont {T.}~\bibnamefont {Mochizuki}}, \bibinfo {author} {\bibfnamefont {C.}~\bibnamefont {Kim}}, \bibinfo {author} {\bibfnamefont {M.}~\bibnamefont {Yoshita}}, \bibinfo {author} {\bibfnamefont {H.}~\bibnamefont {Akiyama}}, \bibinfo {author} {\bibfnamefont {L.~N.}\ \bibnamefont {Pfeiffer}}, \ and\ \bibinfo {author} {\bibfnamefont {K.~W.}\ \bibnamefont {West}},\ }\href {\doibase 10.1063/1.4936753} {\bibfield  {journal} {\bibinfo  {journal} {Appl. Phys. Lett.}\ }\textbf {\bibinfo {volume} {107}} (\bibinfo {year} {2015}),\ 10.1063/1.4936753}\BibitemShut {NoStop}%
\bibitem [{\citenamefont {Hecht}(2017)}]{hecht2017}%
  \BibitemOpen
  \bibfield  {author} {\bibinfo {author} {\bibfnamefont {E.}~\bibnamefont {Hecht}},\ }\href {https://ebookcentral.proquest.com/lib/ethz/reader.action?docID=5833124&ppg=1} {\emph {\bibinfo {title} {Optics, Global Edition}}}\ (\bibinfo  {publisher} {Pearson Higher Education \& Professional Group},\ \bibinfo {year} {2017})\ p.\ \bibinfo {pages} {704}\BibitemShut {NoStop}%
\bibitem [{\citenamefont {Diels}(2006)}]{diels2006}%
  \BibitemOpen
  \bibfield  {author} {\bibinfo {author} {\bibfnamefont {J.-C.}\ \bibnamefont {Diels}},\ }\href@noop {} {\emph {\bibinfo {title} {Ultrashort laser pulse phenomena fundamentals, techniques, and applications on a femtosecond time scale}}},\ \bibinfo {edition} {2nd}\ ed.,\ Optics and photonics\ (\bibinfo  {publisher} {Elsevier/Academic Press},\ \bibinfo {address} {Amsterdam},\ \bibinfo {year} {2006})\BibitemShut {NoStop}%
\bibitem [{\citenamefont {Lu}\ \emph {et~al.}(2014)\citenamefont {Lu}, \citenamefont {Tsou}, \citenamefont {Chen}, \citenamefont {Chen}, \citenamefont {Cheng}, \citenamefont {Yang}, \citenamefont {Chen}, \citenamefont {Hsu},\ and\ \citenamefont {Kung}}]{lu2014}%
  \BibitemOpen
  \bibfield  {author} {\bibinfo {author} {\bibfnamefont {C.-H.}\ \bibnamefont {Lu}}, \bibinfo {author} {\bibfnamefont {Y.-J.}\ \bibnamefont {Tsou}}, \bibinfo {author} {\bibfnamefont {H.-Y.}\ \bibnamefont {Chen}}, \bibinfo {author} {\bibfnamefont {B.-H.}\ \bibnamefont {Chen}}, \bibinfo {author} {\bibfnamefont {Y.-C.}\ \bibnamefont {Cheng}}, \bibinfo {author} {\bibfnamefont {S.-D.}\ \bibnamefont {Yang}}, \bibinfo {author} {\bibfnamefont {M.-C.}\ \bibnamefont {Chen}}, \bibinfo {author} {\bibfnamefont {C.-C.}\ \bibnamefont {Hsu}}, \ and\ \bibinfo {author} {\bibfnamefont {A.~H.}\ \bibnamefont {Kung}},\ }\href {\doibase 10.1364/OPTICA.1.000400} {\bibfield  {journal} {\bibinfo  {journal} {Optica}\ }\textbf {\bibinfo {volume} {1}},\ \bibinfo {pages} {400} (\bibinfo {year} {2014})}\BibitemShut {NoStop}%
\bibitem [{\citenamefont {H\"{a}drich}\ \emph {et~al.}(2016)\citenamefont {H\"{a}drich}, \citenamefont {Kienel}, \citenamefont {M\"{u}ller}, \citenamefont {Klenke}, \citenamefont {Rothhardt}, \citenamefont {Klas}, \citenamefont {Gottschall}, \citenamefont {Eidam}, \citenamefont {Drozdy}, \citenamefont {J\'{o}j\'{a}rt}, \citenamefont {V\'{a}rallyay}, \citenamefont {Cormier}, \citenamefont {Osvay}, \citenamefont {T\"{u}nnermann},\ and\ \citenamefont {Limpert}}]{hadrich2016}%
  \BibitemOpen
  \bibfield  {author} {\bibinfo {author} {\bibfnamefont {S.}~\bibnamefont {H\"{a}drich}}, \bibinfo {author} {\bibfnamefont {M.}~\bibnamefont {Kienel}}, \bibinfo {author} {\bibfnamefont {M.}~\bibnamefont {M\"{u}ller}}, \bibinfo {author} {\bibfnamefont {A.}~\bibnamefont {Klenke}}, \bibinfo {author} {\bibfnamefont {J.}~\bibnamefont {Rothhardt}}, \bibinfo {author} {\bibfnamefont {R.}~\bibnamefont {Klas}}, \bibinfo {author} {\bibfnamefont {T.}~\bibnamefont {Gottschall}}, \bibinfo {author} {\bibfnamefont {T.}~\bibnamefont {Eidam}}, \bibinfo {author} {\bibfnamefont {A.}~\bibnamefont {Drozdy}}, \bibinfo {author} {\bibfnamefont {P.}~\bibnamefont {J\'{o}j\'{a}rt}}, \bibinfo {author} {\bibfnamefont {Z.}~\bibnamefont {V\'{a}rallyay}}, \bibinfo {author} {\bibfnamefont {E.}~\bibnamefont {Cormier}}, \bibinfo {author} {\bibfnamefont {K.}~\bibnamefont {Osvay}}, \bibinfo {author} {\bibfnamefont {A.}~\bibnamefont {T\"{u}nnermann}}, \ and\ \bibinfo {author} {\bibfnamefont {J.}~\bibnamefont {Limpert}},\ }\href {\doibase
  10.1364/OL.41.004332} {\bibfield  {journal} {\bibinfo  {journal} {Opt. Lett.}\ }\textbf {\bibinfo {volume} {41}},\ \bibinfo {pages} {4332} (\bibinfo {year} {2016})}\BibitemShut {NoStop}%
\bibitem [{\citenamefont {Wu}\ and\ \citenamefont {Zhang}(1995)}]{wu1995}%
  \BibitemOpen
  \bibfield  {author} {\bibinfo {author} {\bibfnamefont {Q.}~\bibnamefont {Wu}}\ and\ \bibinfo {author} {\bibfnamefont {X.-C.}\ \bibnamefont {Zhang}},\ }\href {\doibase 10.1063/1.114909} {\bibfield  {journal} {\bibinfo  {journal} {Appl. Phys. Lett.}\ }\textbf {\bibinfo {volume} {67}},\ \bibinfo {pages} {3523} (\bibinfo {year} {1995})}\BibitemShut {NoStop}%
\bibitem [{\citenamefont {Lu}\ \emph {et~al.}(1997)\citenamefont {Lu}, \citenamefont {Campbell},\ and\ \citenamefont {Zhang}}]{Lu1997}%
  \BibitemOpen
  \bibfield  {author} {\bibinfo {author} {\bibfnamefont {Z.~G.}\ \bibnamefont {Lu}}, \bibinfo {author} {\bibfnamefont {P.}~\bibnamefont {Campbell}}, \ and\ \bibinfo {author} {\bibfnamefont {X.-C.}\ \bibnamefont {Zhang}},\ }\href {\doibase 10.1063/1.119803} {\bibfield  {journal} {\bibinfo  {journal} {Appl. Phys. Lett.}\ }\textbf {\bibinfo {volume} {71}},\ \bibinfo {pages} {593} (\bibinfo {year} {1997})}\BibitemShut {NoStop}%
\bibitem [{\citenamefont {Hong}\ \emph {et~al.}(2002)\citenamefont {Hong}, \citenamefont {Kim}, \citenamefont {Kang},\ and\ \citenamefont {Nam}}]{hong2002}%
  \BibitemOpen
  \bibfield  {author} {\bibinfo {author} {\bibfnamefont {K.-H.}\ \bibnamefont {Hong}}, \bibinfo {author} {\bibfnamefont {J.-H.}\ \bibnamefont {Kim}}, \bibinfo {author} {\bibfnamefont {Y.~H.}\ \bibnamefont {Kang}}, \ and\ \bibinfo {author} {\bibfnamefont {C.~H.}\ \bibnamefont {Nam}},\ }\href {\doibase 10.1007/s00340-002-0889-5} {\bibfield  {journal} {\bibinfo  {journal} {Appl. Phys. B}\ }\textbf {\bibinfo {volume} {74}},\ \bibinfo {pages} {s231} (\bibinfo {year} {2002})}\BibitemShut {NoStop}%
\bibitem [{\citenamefont {Jazbin\v{s}ek}\ \emph {et~al.}(2019)\citenamefont {Jazbin\v{s}ek}, \citenamefont {Puc}, \citenamefont {Abina},\ and\ \citenamefont {Zidansek}}]{jazbinsek2019}%
  \BibitemOpen
  \bibfield  {author} {\bibinfo {author} {\bibfnamefont {M.}~\bibnamefont {Jazbin\v{s}ek}}, \bibinfo {author} {\bibfnamefont {U.}~\bibnamefont {Puc}}, \bibinfo {author} {\bibfnamefont {A.}~\bibnamefont {Abina}}, \ and\ \bibinfo {author} {\bibfnamefont {A.}~\bibnamefont {Zidansek}},\ }\href {\doibase 10.3390/app9050882} {\bibfield  {journal} {\bibinfo  {journal} {Applied Sciences}\ }\textbf {\bibinfo {volume} {9}},\ \bibinfo {pages} {882} (\bibinfo {year} {2019})}\BibitemShut {NoStop}%
\bibitem [{\citenamefont {Boyd}(2003)}]{boyd2003}%
  \BibitemOpen
  \bibfield  {author} {\bibinfo {author} {\bibfnamefont {R.~W.}\ \bibnamefont {Boyd}},\ }\href {\doibase 10.1016/B978-0-12-121682-5.X5000-7} {\emph {\bibinfo {title} {Nonlinear {Optics}}}}\ (\bibinfo  {publisher} {Elsevier},\ \bibinfo {year} {2003})\BibitemShut {NoStop}%
\bibitem [{\citenamefont {Cartella}\ \emph {et~al.}(2018)\citenamefont {Cartella}, \citenamefont {Nova}, \citenamefont {Fechner}, \citenamefont {Merlin},\ and\ \citenamefont {Cavalleri}}]{cartella2018}%
  \BibitemOpen
  \bibfield  {author} {\bibinfo {author} {\bibfnamefont {A.}~\bibnamefont {Cartella}}, \bibinfo {author} {\bibfnamefont {T.~F.}\ \bibnamefont {Nova}}, \bibinfo {author} {\bibfnamefont {M.}~\bibnamefont {Fechner}}, \bibinfo {author} {\bibfnamefont {R.}~\bibnamefont {Merlin}}, \ and\ \bibinfo {author} {\bibfnamefont {A.}~\bibnamefont {Cavalleri}},\ }\href {\doibase 10.1073/pnas.1809725115} {\bibfield  {journal} {\bibinfo  {journal} {Proc. Natl. Acad. Sci.}\ }\textbf {\bibinfo {volume} {115}},\ \bibinfo {pages} {12148} (\bibinfo {year} {2018})}\BibitemShut {NoStop}%
\bibitem [{\citenamefont {Henstridge}\ \emph {et~al.}(2022)\citenamefont {Henstridge}, \citenamefont {F\"{o}rst}, \citenamefont {Rowe}, \citenamefont {Fechner},\ and\ \citenamefont {Cavalleri}}]{henstridge2022}%
  \BibitemOpen
  \bibfield  {author} {\bibinfo {author} {\bibfnamefont {M.}~\bibnamefont {Henstridge}}, \bibinfo {author} {\bibfnamefont {M.}~\bibnamefont {F\"{o}rst}}, \bibinfo {author} {\bibfnamefont {E.}~\bibnamefont {Rowe}}, \bibinfo {author} {\bibfnamefont {M.}~\bibnamefont {Fechner}}, \ and\ \bibinfo {author} {\bibfnamefont {A.}~\bibnamefont {Cavalleri}},\ }\href {\doibase 10.1038/s41567-022-01512-3} {\bibfield  {journal} {\bibinfo  {journal} {Nat. Phys.}\ }\textbf {\bibinfo {volume} {18}},\ \bibinfo {pages} {457} (\bibinfo {year} {2022})}\BibitemShut {NoStop}%
\bibitem [{\citenamefont {Basini}\ \emph {et~al.}(2024)\citenamefont {Basini}, \citenamefont {Pancaldi}, \citenamefont {Wehinger}, \citenamefont {Udina}, \citenamefont {Unikandanunni}, \citenamefont {Tadano}, \citenamefont {Hoffmann}, \citenamefont {Balatsky},\ and\ \citenamefont {Bonetti}}]{basini2024}%
  \BibitemOpen
  \bibfield  {author} {\bibinfo {author} {\bibfnamefont {M.}~\bibnamefont {Basini}}, \bibinfo {author} {\bibfnamefont {M.}~\bibnamefont {Pancaldi}}, \bibinfo {author} {\bibfnamefont {B.}~\bibnamefont {Wehinger}}, \bibinfo {author} {\bibfnamefont {M.}~\bibnamefont {Udina}}, \bibinfo {author} {\bibfnamefont {V.}~\bibnamefont {Unikandanunni}}, \bibinfo {author} {\bibfnamefont {T.}~\bibnamefont {Tadano}}, \bibinfo {author} {\bibfnamefont {M.~C.}\ \bibnamefont {Hoffmann}}, \bibinfo {author} {\bibfnamefont {A.~V.}\ \bibnamefont {Balatsky}}, \ and\ \bibinfo {author} {\bibfnamefont {S.}~\bibnamefont {Bonetti}},\ }\href {\doibase 10.1038/s41586-024-07175-9} {\bibfield  {journal} {\bibinfo  {journal} {Nature}\ }\textbf {\bibinfo {volume} {628}},\ \bibinfo {pages} {534} (\bibinfo {year} {2024})}\BibitemShut {NoStop}%
\bibitem [{\citenamefont {Lin}\ \emph {et~al.}(2023)\citenamefont {Lin}, \citenamefont {Xu}, \citenamefont {Chen}, \citenamefont {Guan}, \citenamefont {Yao}, \citenamefont {Zhang}, \citenamefont {Li},\ and\ \citenamefont {Zhu}}]{lin2023}%
  \BibitemOpen
  \bibfield  {author} {\bibinfo {author} {\bibfnamefont {T.}~\bibnamefont {Lin}}, \bibinfo {author} {\bibfnamefont {R.}~\bibnamefont {Xu}}, \bibinfo {author} {\bibfnamefont {X.}~\bibnamefont {Chen}}, \bibinfo {author} {\bibfnamefont {Y.}~\bibnamefont {Guan}}, \bibinfo {author} {\bibfnamefont {M.}~\bibnamefont {Yao}}, \bibinfo {author} {\bibfnamefont {J.}~\bibnamefont {Zhang}}, \bibinfo {author} {\bibfnamefont {X.}~\bibnamefont {Li}}, \ and\ \bibinfo {author} {\bibfnamefont {H.}~\bibnamefont {Zhu}},\ }\href {\doibase 10.1021/acsphotonics.3c00787} {\bibfield  {journal} {\bibinfo  {journal} {ACS Photonics}\ }\textbf {\bibinfo {volume} {11}},\ \bibinfo {pages} {33} (\bibinfo {year} {2023})}\BibitemShut {NoStop}%
\end{thebibliography}%

\end{document}